\documentclass{elsart}

\usepackage{amssymb}

\usepackage[dvips]{graphicx}

\newtheorem{theorem}{Theorem}[section]
\theoremstyle{definition}
\renewcommand{\theequation}{\thesection.\arabic{equation}}
\newcommand{\R}{\mathbb{R}}
\newtheorem{rmrk}{Remark}[section]

\begin{document}

\begin{frontmatter}

\title{Nonlinear option pricing models for illiquid markets: scaling
  properties and explicit solutions}

\author{Ljudmila A. Bordag\corauthref{cor1} }

\address{IDE, Halmstad University, Box 823, 301 18 Halmstad, Sweden}
\ead{Ljudmila.Bordag@ide.hh.se}
\ead[url]{http://www2.hh.se/staff/ljbo/}

\corauth[cor1]{Corresponding author. Ljudmila A. Bordag, IDE,
Halmstad University, Box 823, 301 18 Halmstad, Sweden, Fax: +46 35 12 03 48}

\author{R\"udiger Frey}
\address{Department of Mathematics,
Universit\"at Leipzig, Postfach 10 09 20, 04009 Leipzig, Germany}
\ead{ruediger.frey@math.uni-leipzig.de}
\ead[url]{www.math.uni-leipzig.de/~frey}

\begin{abstract}
Several models for the pricing of derivative securities in illiquid markets are
discussed. A typical type of nonlinear  partial differential equations arising from
these investigation is studied. The scaling properties of these equations  are
discussed. Explicit solutions for one of the models are obtained and studied.
\end{abstract}



\end{frontmatter}

\section{Introduction}

Standard derivative pricing theory is based on the assumption of frictionless
and perfectly liquid markets. In particular, it is assumed that all investors
are \emph{small} relative to the market so that they can buy arbitrarily large
quantities of the underlying assets without affecting its price (perfectly
liquid markets). Given the scale of hedging activities on many financial
markets this is clearly unrealistic.  Hence in recent years a number of
approaches for dealing with market illiquidity in the pricing and the hedging
of derivative securities have been developed.  The financial framework that is
being used differs substantially between model classes.  However, as shown in
Section~\ref{sec:illiquid.-markets} below, in all models derivative prices can
be characterized by fully nonlinear versions of the standard parabolic
Black-Scholes PDE. Moreover, these nonlinear Black-Scholes equations have a
very similar structure. This makes these nonlinear Black-Scholes equations a
useful reference point for studying derivative asset analysis under market
illiquidity. In Sections~3 we therefore study scaling properties and compute
explicit solutions for a typical fully nonlinear Black-Scholes equation;
Section~4 is devoted to studying properties of solutions and sensitivities
with respect to model parameters. The relevant literature is discussed in the
body of the paper.

\section{Illiquid markets and nonlinear Black-Scholes equations}
\label{sec:illiquid.-markets}

In order to motivate the subsequent analysis we present a brief synopsis of three
different frameworks for modeling illiquid markets. We group them under the labels
a)\emph{transaction-cost models}; b) \emph{reduced-form SDE-models}; c)
\emph{reaction-function} or \emph{equilibrium models}. In particular, we show that the
value function of a certain type of selffinancing strategies (so called
\emph{Markovian} strategies) must be a solution of a fully nonlinear version of the
standard Black-Scholes equation. In all models there will be two assets, a risk-free
money-market account $B$ which is perfectly liquid and a risky and illiquid asset $S$
(the stock). Without loss of generality, we use the money market account as numeraire;
hence $B_t \equiv 1$, and interest rates can be taken equal to zero.

{\bf (Quadratic) transaction-cost models.} The predominant model in this class has
been put forward by Cetin, Jarrow and Protter \cite{bib:cetin-jarrow-protter-04}. In
this model there is a \emph{fundamental} stock price process $S^{0}$ following
geometric Brownian motion,
\begin{equation}\label{eq:BS-model}
dS_t^0 = \mu S_t^0 dt + \sigma S_t^0 dW_t
\end{equation}
for constants $\mu \in \R $, $\sigma >0$ and a standard Brownian motion
$W$. The \emph{transaction price} to be paid at time $t$ for trading $\alpha$
shares is
\begin{equation} \label{eq:transaction-price}
\bar{S}_t (\alpha) = e^{\rho \alpha} S_t^0, \quad \alpha \in \R,\,\,\,\rho > 0,
\end{equation}
where $\rho$ is a liquidity parameter.

Intuitively, in the model (\ref{eq:transaction-price}) a trader has to pay a spread
whose size relative to the fundamental price, $S_t^0(e^{\rho \alpha} -1)$, depends on
the amount $\alpha$ which is traded. As shown by \cite{bib:cetin-jarrow-protter-04},
this leads to transaction costs which are proportional to the \emph{quadratic
variation} of the stock trading strategy. In order to formulate this statement more
precisely we consider a selffinancing trading strategy $(\Phi_t, \eta_t)_{t\ge 0}$
giving the number of stocks and the position in the money market for predictable
stochastic processes $\Phi$ and $\eta$. The \emph{value} of this strategy at time $t$
equals $V_t = \Phi_t S_t^0 + \eta_t$.


Consider now a \emph{Markovian strategy}, i.e. a trading strategy of the form
$\Phi_t = \phi(t, S_t^0)$ for a smooth function $\phi$.  In this case $\Phi$
is a semimartingale with quadratic variation given by
$$[ \Phi]_t =\int_0^t \left (\phi_S(s,S^0_s) \sigma S_s^0 \right)^2 ds \,;$$
see for instance Chapter~2 of \cite{bib:protter-92}. Then Theorem A3 of
\cite{bib:cetin-jarrow-protter-04} yields the following dynamics of $V_t$
\begin{equation} \label{eq:Vt-protter}
 d V_t = \phi(t,S_t^0) dS_t^0 - \rho S_t^0 d [ \Phi]_t =
         \phi(t,S_t^0) dS_t^0 - \rho S_t^0 \left ( \phi_S(t,S^0_t) \sigma S_t^0\right )^2 dt.
\end{equation}
Note that according to (\ref{eq:Vt-protter}), the market impact of the large
trader leads to an extra transaction-cost term of size $\rho \int_0^t S_s^0 \,
d [ \Phi]_s $ in the wealth dynamics.

Suppose now that $u$ and $\phi$ are smooth functions such that $u(t,S_t^0)$
gives the value of a selffinancing trading strategy with stock position
$\phi(t,S_t^0)$.  According to the It\^o formula, the process
$(u(t,S_t^0))_{t\ge 0}$ has dynamics
$$ d u(t,S_t^0) = u_S(t,S_t^0) dS_t^0 + \left ( u_t (t,S_t^0) + \frac{1}{2} \sigma ^2 (S_t^0)^2
   u_{SS}(t,S_t^0)\right ) \, dt \,.
$$ Comparing this with (\ref{eq:Vt-protter}), using the uniqueness of
semimartingale decompositions it is immediate that $u$ must satisfy the PDE
$u_t + \frac{1}{2} \sigma^2 S^2 u_{SS} + \rho S^3 \sigma^2 \phi_S =0$ and that
moreover $\phi =u_S$. The last relation gives $\phi_S =u_{SS}$, so that we
obtain the following nonlinear PDE for $u$
\begin{equation}
\label{eq:nonlinear-BS-protter}
 u_t  + \frac{1}{2} \sigma^2 S^2 u_{SS} \left (1 + 2 \rho S u_{SS}\right ) =0\,.
  \end{equation}
When pricing a derivative security with maturity date $T$ and payoff $h(S_T)$ for some
function $h: [0,\infty) \to \R$ we have to add the terminal condition $u(T,S) = h(S)$,
$S\ge 0$. For instance, in case of a European call option with strike price $K > 0$ we
have $h(S) = \max\{S-K,0\}$.

 Note that the original paper
\cite{bib:cetin-jarrow-protter-04} goes further in the analysis of quadratic
transaction cost models. To begin with, a general framework is proposed that contains
(\ref{eq:transaction-price}) as special (but typical) case. Moreover, conditions for
absence of arbitrage for a general class of trading strategies - containing Markovian
trading strategies as special case - are given, and a notion of approximative market
completeness is studied.

{\bf Reduced-form SDE models.} Under this modeling approach it is assumed that
investors are \emph{large traders} in the sense that their trading activity affects
equilibrium stock prices. More precisely, given a liquidity parameter $\rho \ge 0$ and
a semimartingale $\Phi$ representing the stock trading strategy of a given trader, it
is assumed that the stock price satisfies the SDE
\begin{equation}
\label{eq:price-SDE-model} dS_t = \sigma S_{t}d W_t + \rho  S_{t} d\Phi_t\,.
\end{equation}
The intuitive interpretation is as follows given that the  investor buys (sells) stock
($\Delta \Phi_t >0$) the stock price is pushed (downward) upward by $\rho S_{t-}
\Delta \Phi_t$; the strength of this price impact depends on the parameter $\rho$.
Note that for $\rho=0$ the asset price simply follows a Black-Scholes model with
reference volatility $\sigma$. The model (\ref{eq:price-SDE-model}) is studied among
others in \cite{bib:frey-00c}, \cite{bib:frey-patie-02},
 \cite{JandarckaSevcovic} or \cite{bib:liu-yong-05}.

In the sequel we denote the asset price process which results if a large  uses a
particular trading strategy $\Phi$ by $S^{\Phi}$. Suppose as before that the trading
strategy is Markovian, i.e.~of the form $\Phi_t = \phi(t, S_t)$ for a smooth function
$\phi$. Applying the It\^o formula to (\ref{eq:price-SDE-model}) shows that $S^{\phi}$
is an It\^o process with dynamics
$$d S_t^\phi = v^{\phi}(t,S_t^\phi)S_t^ \phi dW_t
+b^{\phi}(t,S_t^\phi)S_t^\phi dt$$ with \emph{adjusted volatility} given by
\begin{equation} \label{eq:adjusted-vol-1}
 v^{\phi}(t,S) = \frac{\sigma}{1 - \rho  S \phi_S(t, S)}\, ;
\end{equation}
see for instance \cite{bib:frey-00c} for details. In the
model~(\ref{eq:price-SDE-model}), a portfolio with stock trading strategy $\Phi$ and
value $V$ is termed \emph{selffinancing}, if satisfies the equation $dV_t = \Phi_t
dS_t^\Phi$. Note that the form of the strategy $\Phi$ affects the dynamics of $S$;
this \emph{feedback effect} will give rise to nonlinearities in the wealth dynamics as
we now show. Suppose that $V_t = u(t,S_t)$ and $\Phi_t = \phi(t,S_t)$ for smooth
functions $u$ and $\phi$. As before, applying the It\^o formula to the process
$(u(t,S_t^\phi))_{t\ge 0}$ yields $\phi_t = u_S $. Moreover, $u$ must satisfy the
relation $u_t + \frac{1}{2} (v^{\phi})^2(t,S) S^2 u_{SS} =0$. Using
(\ref{eq:adjusted-vol-1}) and the relation $\phi_S = u_{SS}$ we thus obtain the
following fully nonlinear PDE for $u(t,S)$
\begin{equation}\label{eq:nonlinear-BS-SDE} u_t + \frac{1}{2} \frac{\sigma^2 }{(1-\rho
S u_{SS})^2} S^2 u_{SS} =0\,.
\end{equation}
Again, for pricing derivative securities a terminal condition corresponding to the
particular payoff at hand needs to be added.

{\bf Equilibrium or reaction-function models.} Here the model primitive is a smooth
\emph{reaction function} $\psi$ that gives the equilibrium stock price $S_t$ at time
$t$ as function of some fundamental value $F_t$ and the stock position of  a large
trader. A reaction function can be seen as reduced-form representation of an economic
equilibrium model, such as the models proposed in \cite{bib:frey-stremme-97},
\cite{bib:platen-schweizer-98} or \cite{bib:schonbucher-wilmott-00}.  In these models
there are two types of traders in the market: ordinary investors and a large investor.
The overall supply of the stock is normalized to one. The normalized stock demand of
the ordinary investors at time $t$ is modelled as a function $D(F_t,x)$ where $x$ is
the proposed price of the stock. The normalized stock demand of the large investor is
written the form $\rho \Phi_t$; $\rho \ge 0 $ is a parameter that measures the size of
the trader's position relative to the total supply of the stock. The equilibrium price
$S_t$ is then determined by the market clearing condition
\begin{equation}
  \label{eq:equilibrium}
  D(F_t,S_t) + \rho \Phi_t =1\,.
\end{equation}
Under suitable assumptions on $D$ equation (\ref{eq:equilibrium}) admits a
unique solution, hence $S_t$ can be expressed as a function $\psi$ of $F_t$
and $\rho \Phi_t$, i.e. $S_t = \psi(F_t,\rho \Phi_t)$. For instance we have in
\cite{bib:platen-schweizer-98} $\psi(f, \alpha)= f \exp(\alpha ) $; the model
used in \cite{bib:frey-stremme-97} and \cite{bib:papanicolaou-sircar-98} leads
to the reaction function $\psi(f,\alpha)= f/(1-\alpha)\,.$ The
reaction-function approach is also used in \cite{bib:jarrow-94} and in
\cite{bib:frey-98a}.

Now we turn to the characterization of selffinancing hedging strategies in
reaction-function models. Throughout we assume that the fundamental-value process $F$
follows a geometric Brownian motion with volatility $\sigma$ (see
(\ref{eq:BS-model})). Moreover, we assume that the reaction function is of the form
$\psi(f, \alpha) = f g(\alpha)$ for some increasing function $g$. This holds for the
specific examples introduced above and, more generally, for any model where $D(f,x) =
U(f/x)$ for a strictly increasing function $U \colon (0,\infty) \to \R $ with suitable
range.

Assuming as before that the normalized trading strategy of the large trader is of the
form $\rho \phi(t,S)$ for a smooth function $\phi$ we get from It\^o's formula, since
$S_t = g(\rho \phi(t, S_t))F_t$,
 $$ dS_t = g(\rho \phi(t, S_t))\, dF_t + \rho F_t g_\alpha(\rho \phi(t, S_t))
 \phi_S(t, S_t)\,d S_t + b (t,S_t)\, dt ;$$ the precise form of $b(t,S_t) $ is
 irrelevant for our purposes. Rearranging and integrating the expression
 $\big(1 - \rho F_t g_\alpha(\rho \phi(t, S_t)) \phi_S(t, S_t)\big)^{-1}$ over
 both sides of this equation gives the following dynamics of $S$:
\begin{equation} \label{eq:price-reaction-function-model}
dS_t = \frac{1}{1- \rho \frac{g_\alpha (\rho \phi(t,S_t))}{g(\rho
\phi(t,S_t))} S_t \phi_S(t,S_t) } \sigma S_t dW_t + \tilde{b}(t,S_t) dt \,.
\end{equation}
\begin{rmrk} A very general analysis of the dynamics of selffinancing strategies in reaction-function
models (and generalizations thereof) using the \emph{It{\^o}-Wentzell formula} can be
found in \cite{bib:bank-baum-04}.
\end{rmrk}
A similar reasoning as in the case of the reduced-form SDE models now gives
the following PDE for the value function $u(t,S)$ of a selffinancing strategy
\begin{equation} \label{eq:nonlinear-BS-psi}
 u_t + \frac{1}{2} \frac{\sigma^2 }{\left(1-\rho \frac{g_\alpha (\rho u_{S})}{g(\rho u_{S})}
   S u_{SS}\right)^2} S^2 u_{SS}  =0\,.
\end{equation}
In particular, for $g(\alpha) = \exp(\alpha)$ we have $g = g_\alpha$ and
(\ref{eq:nonlinear-BS-psi}) reduces to equation (\ref{eq:nonlinear-BS-SDE});
for $g(\alpha) = {1} \big /{(1- \alpha)}$ as in \cite{bib:frey-stremme-97},
\cite{bib:papanicolaou-sircar-98}, we get the PDE
\begin{equation} \label{eq:nonlinear-BS-sircar}
 u_t + \frac{1}{2} \frac{\sigma^2(1-\rho u_S)^2 }{\left(1-\rho u_S - \rho
   Su_{SS}\right)^2} S^2 u_{SS}  =0 \,.
\end{equation}

{\bf Nonlinear Black-Scholes equations.}  The nonlinear PDEs
(\ref{eq:nonlinear-BS-protter}), (\ref{eq:nonlinear-BS-SDE}) ,
(\ref{eq:nonlinear-BS-psi}) and
(\ref{eq:nonlinear-BS-sircar}) are all of the form
\begin{equation}\label{eq:nonlinear-BS-general}
 u_t + \frac{1}{2} \sigma^2 v(S,\rho u_S, \rho u_{SS}) S^2 u_{SS} =0 \,,
\end{equation}
where $v(S,0,0)=1$. Since $\rho$ is often considered to be small, it is of
interest to replace $v$ with its first order Taylor approximation around $\rho
=0$. It is immediately seen that for the equations (\ref{eq:nonlinear-BS-SDE})
and (\ref{eq:nonlinear-BS-sircar}) this linearization is given by ${v}(S,\rho
u_S, \rho u_{u_SS}) \approx 1 + \rho S u_{SS};$ replacing $v$ with this first
order Taylor approximation in (\ref{eq:nonlinear-BS-SDE}) and
(\ref{eq:nonlinear-BS-sircar}) thus immediately leads to the PDE
(\ref{eq:nonlinear-BS-protter}).

Note that (\ref{eq:nonlinear-BS-general}) is a fully nonlinear equation in the
sense that the coefficient of the highest derivative is a nonlinear function
of this derivative. A similar feature can be observed for the limiting price
in certain transaction cost models under a proper rescaling of transaction
cost and trading frequency; see for instance
\cite{bib:barles-soner-98}. Nonlinear PDEs for incomplete markets obtained via
exponential utility indifference hedging such as \cite{bib:becherer-04} or
\cite{bib:musiela-zariphopoulou-04} on the other hand are quasi-linear
equations in the sense that the highest derivative $u_{SS}$ enters the
equation in a linear way, similar to the well-known reaction-diffusion
equations arising in physics or chemistry.  From an analytical point of view
the nonlinearities arising in both cases are quite different. In the latter
quasi-linear case we have to do with a regular perturbation of the classical
Black-Scholes (BS) equation but in the case (\ref{eq:nonlinear-BS-psi}) we have
to do with singular perturbation, meaning that the highest derivative is
included in the perturbation \cite{Johnson}.  In case of a regular
perturbation it is typical to look for a representation of a solutions to a
quasi-linear equation in the form
\begin{equation}
u\left(S,t\right)=u_{BS}(S,t) + \sum_{n=1}^{N}\rho u_1(S,t) +O(\rho^{N+1}). \label{as}
\end{equation}
For many forms of nonlinearities the uniform convergence as $\rho \to 0 $ of
the expansions of type (\ref{as}) can be established. If the perturbation is
of the singular type, the asymptotic expansion of type (\ref{as}) typically
breaks down for some $S,t$ and some $N\ge 0$. It can as well happens that the
definition domain for the solutions of type (\ref{as}) include just one point
or that it is empty. It is very important to obtain explicit solutions for
such models because in these case we can not hope to get good approximations
for solutions by expansions of the type (\ref{as}) in the whole region.

\section{Invariant Solutions for a Nonlinear Black-Scholes Equation}

Our goal in this section is to investigate the nonlinear Black-Scholes equation
(\ref{eq:nonlinear-BS-SDE}) using analytical methods. As we have just seen, this
equation is typical for the nonlinearities arising in pricing equations for
derivatives in illiquid markets in general.\footnote{Our approach can be applied to
equations~(\ref{eq:nonlinear-BS-protter}) and (\ref{eq:nonlinear-BS-sircar}) as well;
see \cite{Bobrov, Sergeeva} for details.} Using the symmetry group and its invariants
the partial differential equation (\ref{eq:nonlinear-BS-SDE}) can be reduced in
special cases to ordinary differential equations. In \cite{Bordag:2004} a special
family of invariant solutions to the equation (\ref{eq:nonlinear-BS-SDE}) was studied;
in particular, the explicit solutions were used as test case for various numerical
methods. In the present paper we study \emph{all} invariant solutions to equation
(\ref{eq:nonlinear-BS-SDE}).

\subsection{General Results}

In order to describe the symmetry group of (\ref{eq:nonlinear-BS-SDE}) we use results
from Bordag ~\cite{Bordag:2005}. In that paper the slightly more general equation
\begin{eqnarray} \label{ur2lam}
u_t+\frac{\sigma^2 S^2}2\frac{u_{SS}}{(1-\rho \lambda (S) S u_{SS})^2}=0,
\end{eqnarray}
with a continuous function $\lambda: (0,\infty) \to (0,\infty)$ is studied
(note that for $\lambda \equiv 1$ this equation reduces to
(\ref{eq:nonlinear-BS-SDE})) and the following theorem was proved.

\begin{theorem}
\label{symteor}
The action of the symmetry group $G_{\Delta}$ of (\ref{ur2lam}) with an
  arbitrary function $\lambda(S)$ is given by
\begin{eqnarray}
{\tilde S}&=&S,\label{str}\\
{\tilde t}&=&t+ a_2 \epsilon,\label{ttr}\\
{\tilde u}&=&u +a_3 S \epsilon +a_4 \epsilon,~~ \epsilon \in (- \infty,\infty). \label{utro}
\end{eqnarray}
If the function $\lambda(S)$ has the special form $ \lambda(S)= \omega S^k,~ k,\omega
\in {\R} $, equation~(\ref{ur2lam}) takes the special form
\begin{equation}\label{uravlam}
u_t+\frac{\sigma^2 S^2}2\frac{u_{SS}}{(1-b S^{k+1} u_{SS})^2}=0\,,b=\omega \rho .
\end{equation}
In that case the  symmetry group $G_{\Delta}$ has the richer structure
\begin{eqnarray}
{\tilde S}&=&S e^{a_1 \epsilon},~~ \epsilon \in (- \infty,\infty ),\label{strsp}\\
{\tilde t}&=&t+ a_2 \epsilon,\label{ttrsp}\nonumber\\
{\tilde u}&=&u e^{a_1 (1-k) \epsilon} + \frac{a_3}{a_1 k} S \epsilon e^{a_1 \epsilon} (1-e^{- a_1 k\epsilon} )   \nonumber\\
&+&\frac{a_4}{a_1(1-k)}(e^{a_1 (1-k) \epsilon} -1),~~ k \ne 0,~k\ne 1
\label{utrnesp}\\
{\tilde u}&=&u e^{a_1\epsilon} + a_3 S \epsilon e^{a_1 \epsilon} +
\frac{a_4}{a_1}(e^{a_1 \epsilon} -1), ~k= 0, \nonumber\\
{\tilde u}&=&u  + \frac{a_3}{a_1} S( e^{a_1 \epsilon} -1) +
a_4 \epsilon, ~k= 1. \label{utrosp}
\end{eqnarray}
\end{theorem}
The proof is based on methods of Lie point symmetries, i.e.~the Lie symmetry algebras
and groups to the corresponding equations were found; see \cite{Bordag:2005} for
details and \cite{Lie, Olver, Ibragimov} for a general introduction to the
methodology. As we will see later, the solutions found in \cite{Bordag:2005} for the
case $\lambda(S)=S$ can be used to obtain the complete set of invariant solutions to
equation (\ref{eq:nonlinear-BS-SDE}).

For $ \lambda(S)=\omega S^k$ with $k=0,1 $  we can use (\ref{strsp}) to obtain a new
independent invariant variable $z$ and (\ref{utrosp}) to obtain a new dependent
variable $v$; these variables are given by
\begin{eqnarray}
z&=&\log S +a t, ~~ a \ne 0,\nonumber\\
v&=& u~ S^{(k-1)},~~ k=0,1. \label{newvar}
\end{eqnarray}
Using these variables we can reduce the partial differential equation (\ref{uravlam})
to the ordinary differential equation
\begin{equation}
v_z + q \frac{v_{zz}  + \xi v_z } {(1- b (v_{zz}+ \xi v_z ))^2}=0,  \label{speck1}\,
\end{equation}
where $ q = \frac{\sigma^2}{2 a},~ a \ne 0, b= \omega \rho \ne 0, \xi=(-1)^k, ~k=0,1.
$ It is a straightforward consequence of Theorem~\ref{symteor} that these are the only
nontrivial invariant variables. In the sequel we will determine explicit solutions for
(\ref{speck1}) for the case $k=0$ which corresponds to the original equation
(\ref{eq:nonlinear-BS-SDE}).

\begin{rmrk} Equation (\ref{speck1}) with $k=0$ and $k=1$ are related
to each other. Note however, that the relation between the
corresponding solutions is not so straightforward, because (\ref{speck1}) is
nonlinear and we need real valued solutions. Hence the results from \cite{Bordag:2005} - where (\ref{speck1})
was studied for the case $k=1$ - do not carry over directly to the present
case $k=0$, so that a detailed analysis of the case $k=0$ is necessary.
\end{rmrk}

To find  families of invariant solution  we introduce a new dependent variable
\begin{equation}
y(z)=v_z(z). \label{samy}
\end{equation}
 If we assume that the denominator of the
equation (\ref{speck1}) is different from zero, we can multiply both terms of
equation (\ref{speck1}) by the denominator of the second term and obtain
\begin{equation}
  y y_z^2 + 2 \xi \left(y^2 - \xi \frac{1}{b}y + \xi \frac{q}{2 b^2} \right)
 y_z + \left(y^2 - \xi \frac{2}{b}y +\left(\frac{1+ \xi q}{ b^2}
 \right)\right) y =0,~~b \ne 0.\label{uryk1}
\end{equation}
We denote the left hand side of this equation by $ F(y,y_z)$.  The equation
(\ref{uryk1}) can possess exceptional solutions which are the solutions of a
system
\begin{equation}
\frac{\partial F(y,y_z)}{\partial y_z}=0,~~ F(y,y_z)=0. \label{exepsys}
\end{equation}
The first equation in this system defines a discriminant
curve which has the form
\begin{equation}
y(z)= \frac{q}{4 b}. \label{discr}
\end{equation}
If this curve is also a solution of the original equation (\ref{uryk1}) then we
obtain an
exceptional solution to equation (\ref{uryk1}).
We obtain an exceptional solution if
$q=- 4 \xi  $, i.e. $a=\xi \sigma^2/8 $. It has the form
\begin{equation}
y(z)= - \frac{\xi}{b},~~ \xi=(-1)^k, k=0,1. \label{exesol}
\end{equation}
This solution belongs to the family of solutions (\ref{yconst}) by the
specified value of the parameter $q$. In all other cases the equation
(\ref{uryk1}) does not possess any exceptional solutions.

Hence the set of solutions of equation (\ref{uryk1}) is the union of the
solution sets of following equations
\begin{eqnarray}
y&=&0,\label{ynul}\\
y&=&\left(\xi \pm \sqrt{q}\right)/b,\label{yconst}\\
y_z&=&\left(- \xi y^2 + \frac{1}{b}y   -\frac{q}{2 b^2} - \sqrt{
 \frac{q}{ b^3} \left( \frac{q}{4 b}
-y\right)}\right) \frac{1}{y},~ y\ne0 \label{yzeqm}\\
y_z&=&\left(-\xi y^2 + \frac{1}{b}y   -\frac{q}{2 b^2} + \sqrt{
 \frac{q}{ b^3} \left( \frac{q}{4 b}
-y\right)}\right) \frac{1}{y}, ~ y\ne0 \label{yzeqp}
\end{eqnarray}
where one of the solutions (\ref{yconst}) is an exceptional solution
(\ref{exesol}) by $q=- 4 \xi $ for $k=1$. In case $k=0$ we have
$\xi =1$ so that solutions of (\ref{yconst}) are complex valued functions. We
denote the right hand side of equations (\ref{yzeqm}), (\ref{yzeqp}) by
$f(y)$.  The Lipschitz condition for equations of the type $y_z=f(y)$ is
satisfied in all points where the derivative $\frac{\partial f}{\partial y}$
exists and is bounded. It is easy to see that this condition will not be
satisfied by
\begin{equation}
y=0,~~ y= \frac{q}{4 b},~~ y=\infty. \label{edin}
\end{equation}
Hence  on the lines (\ref{edin}) the uniqueness of solutions of equations
(\ref{yzeqm}), (\ref{yzeqp}) can be lost. We will study in detail the behavior of
solutions in the neighborhood of the lines (\ref{edin}). For this purpose we look at
the equation (\ref{uryk1}) from another point of view.  If we assume now that
$z,y,y_z$ are complex variables and denote
\begin{equation}
y(z)=\zeta, ~~ y_z(z)=w,~~ \zeta,w \in C, \label{complexsub}
\end{equation}
then the equation (\ref{uryk1}) takes the form
\begin{equation}
F(\zeta,w)= \zeta w^2 + 2 \xi \left(\zeta^2 - \xi\frac{1}{b}\zeta +\xi
 \frac{q}{2 b^2} \right) w + \left(\zeta^2 - \xi \frac{2}{b}\zeta +\frac{1+
 \xi q}{b^2}\right) \zeta =0,\label{vweq}
\end{equation}
where $ b \ne 0.$ The equation (\ref{vweq}) is an algebraic relation in $C^2$
and defines a plane curve in this space.  The polynomial $F(\zeta,w)$ is an
irreducible polynomial if at all roots $w_r(z)$ of $F(\zeta,w_r)$ either the
partial derivative $F_\zeta(\zeta,w_r)$ or $F_w(\zeta,w_r)$ are non equal to
zero.  It is easy to prove that the polynomial (\ref{vweq}) is irreducible.

We can treat equation (\ref{vweq}) as an algebraic relation which defines a
Riemannian surface $\Gamma ~:~ F(\zeta,w)=0~$ of $w=w(\zeta)$ as a compact
manifold over the $\zeta$-sphere.  The function $w(\zeta)$ is uniquely
analytically extended over the Riemann surface $\Gamma$ of two sheets over the
$\zeta-$sphere.  We find all singular or branch points of $w(\zeta)$ if we
study the roots of the first coefficient of the polynomial $F(\zeta,w)$, the
common roots of equations
\begin{equation}
F(\zeta,w)=0,~~F_w(\zeta,w)=0,~~~ \zeta, w \in C\cup{\infty}.
\end{equation}
and the point $\zeta=\infty.$
The set of  singular or branch points consists of the points
\begin{equation}
\zeta_1=0,~~\zeta_2= \frac{q}{4 b}, ~~\zeta_3={\infty} \label{singpoint}.
\end{equation}
As expected we got the same set of points as in real case (\ref{edin}) by the
study of the Lipschitz condition but now the behavior of solutions at the
points is more visible.\\

The points $\zeta_2,\zeta_3$ are the branch points at which two sheets
of $\Gamma$ are glued on. We remark that
 \begin{equation}
w(\zeta_2)=-\frac{1}{b}\left(1-\xi \frac{q}{4} \right) - t \frac{4
\xi}{\sqrt{-bq}}+\cdots,~~ t^2=\zeta-\frac{q}{4 b}, \label{wz2}
\end{equation}
where $t$ is a local parameter in the neighborhood of $\zeta_2.$ For the
special value of $q=4 \xi$, i.e. k=0, the value $w(\zeta_2)$ is equal to zero.

At the point $\zeta_3=\infty$ we have
$$w(\zeta)= - \frac{\xi}{t^2} +\frac{1}{b} - \xi t \sqrt{ \frac{-q}{4 b^3}}, ~~ t^2
=\frac{1}{\zeta},~~ \zeta \to \infty,$$
where $t$ is a local parameter in the neighborhood of $\zeta_3.$
At the point $\zeta_1=0$ the function $w(\zeta)$ has the following behavior
\begin{eqnarray}
w(\zeta)&\sim& -\frac{q}{ b^2}\frac{1}{\zeta},~\zeta \to \zeta_1=0,~ {\mbox{on
the principal sheet}}, \label{pol}\\ w(\zeta) &\sim& -\left(\xi+\frac{1}{q}
\right) \zeta,~\zeta \to \zeta_1=0,~ q \ne 1,~~{\mbox{on the second sheet}},
\label{nul}\\ w(\zeta) &\sim& - 2 b^2 \zeta^2,~\zeta \to \zeta_1=0,~
q=1,~~{\mbox{on the second sheet}}. \label{nul2}
\end{eqnarray}

Any solution $w(\zeta)$ of an irreducible algebraic equation (\ref{vweq}) is
meromorphic on this compact Riemann surface $\Gamma$ of the genus 0 and has a
pole of the order one correspondingly (\ref{pol}) over the point $\zeta_1=0$
and the pole of the second order over $\zeta_3=\infty$.  It means also that
the meromorphic function $w(\zeta)$ cannot be defined on a manifold of less
than 2 sheets over the $\zeta$ sphere.

To solve differential equations (\ref{yzeqm}) and (\ref{yzeqp}) from this
point of view it is equivalent to integrate on $\Gamma$ a differential of the
type $ ~{\frac{{\rm d} \zeta}{w(\zeta)}}~$ and then to solve an Abel's inverse
problem of degenerated type
\begin{equation}
\int{\frac{{\rm d} \zeta}{w(\zeta)}}=z+{\rm const.}  \label{intri}
\end{equation}
The integration can be done very easily because we can introduce a
uniformizing parameter on the Riemann surface $\Gamma$ and represent
the integral (\ref{intri}) in terms of rational functions merged
possibly with logarithmic terms.

To realize this program we
introduce a new variable (our uniformizing parameter $p$) in the way
\begin{eqnarray}
\zeta &=&\frac{q(1-p^2)}{4b} , \label{uniparv}\\
w&=&\frac{\xi (p-1)(q(1+p)^2 +4 \xi)}{4 b  (p+1)}.
\end{eqnarray}
Then the equations (\ref{yzeqm}) and (\ref{yzeqp}) will take the form
\begin{eqnarray}
2 q \xi \int{\frac{p (p+1) {\rm d}p}{(p-1)(q (p+1)^2+4 \xi)}}=z+{\rm const},
\label{privedeqm}\\
2 q \xi \int{\frac{p (p-1) {\rm d}p}{(p+1)(q (p-1)^2+4
\xi)}}=z+{\rm const}\label{privedeqp}.
\end{eqnarray}
The integration procedure of equations (\ref{privedeqm}),
(\ref{privedeqp}) gives rise to
relations defining a complete set of first order differential equations. In
order to see that these are first order ordinary differential equations recall
that from the substitutions (\ref{complexsub}) and (\ref{samy}) we have
\begin{equation}
p=\sqrt{ 1 -\frac{4 b}{q }  v_z }. \label{vzp}
\end{equation}

We summarize all these results in the following  theorem.

\begin{theorem}\label{theoremq}
The equation (\ref{speck1}) for arbitrary values of the
parameters $q, b\ne 0$
can be reduced to the set of first order differential equations which consists
of the equations
\begin{equation}
 v_z=0,~~v_z=(1 \pm 2)/b ,   \label{trivisol}
\end{equation}
and equations (\ref{yzeqm}), (\ref{yzeqp}) where $y$ is defined by the
substitution (\ref{samy}).  The complete set of solutions of the equation
(\ref{speck1}) coincides with the union of solutions of these equations.
\end{theorem}

To solve equations (\ref{privedeqm}), (\ref{privedeqp}) exactly we have
first to integrate and then invert these formulas in order to obtain an exact
representation of $p$ as a function of $z$. If an exact formula for the
function $p=p(z)$ is found we can use the substitution (\ref{vzp}) to obtain
an explicit ordinary differential equation of the type $v_z(z)=f(z)$ or
another suitable type; in that case it is possible to solve the problem
completely.  However, for an arbitrary value of the parameter $q$ inversion is
impossible, and we have just an implicit representation for the solutions of
the equation (\ref{speck1}) as solutions of the implicit first order
differential equations.

\subsection{Exact invariant solutions for
 a fixed relation between $S$ and $t$}

For a special value of the parameter $q$, namely for $q=-4$, we can integrate and
invert the equations (\ref{privedeqm}) and (\ref{privedeqp}). For $q=-4$ the relation
between the variables $S$ and $t$ is fixed in the form
\begin{equation}
z=\log S -\frac{\sigma^2}{8}t,  \label{podszotst}
\end{equation}
equation (\ref{privedeqm}) takes the form
\begin{equation}
(p-1)^2(p+2)=2 ~c~ \exp{(-  3 z/2)} \,,\label{cubm}
\end{equation}
and correspondingly the equation (\ref{privedeqp}) becomes
\begin{equation}
(p+1)^2(p-2)=2  ~c~ \exp{(-3 z/2)}, \label{cubp}
\end{equation}
where $c$ an arbitrary constant. It is easy to see that the equations (\ref{cubm})
and (\ref{cubp}) are connected by a transformation
\begin{equation}
p\to -p, ~~ c\to -c. \label{involut}
\end{equation}
This  symmetry arises from the symmetry of the underlining  Riemann surface
$\Gamma$ (\ref{vweq})
and corresponds to a change of the sheets on  $\Gamma$.
Using these symmetry properties
we can prove following theorem.

\begin{theorem}\label{theorforv}
The second order differential equation
\begin{equation}
v_z - 4 \frac{v_{zz} +v_z }
{(1-b(v_{zz}+v_z ))^2}=0,  \label{speck1q4}
\end{equation}
is exactly integrable for any value of the parameter $b$. The complete set of
solutions for $b \ne 0$ is given by the union of solutions (\ref{solvzplus}),
(\ref{resh1}) -(\ref{resh32}) and solutions
\begin{equation}
v(z)=d,~~v(z)=\frac{3}{b} z+d,~~v(z)=-\frac{1}{b} z+d,\label{trivisol1}
\end{equation}
where $d$ is an arbitrary constant. The last solution in (\ref{trivisol1})
corresponds to the exceptional solution of equation (\ref{uryk1}).

 For $b=0$ equation (\ref{speck1q4}) is linear and its solutions are given by
$v(z)=d_1 + d_2 \exp{-(3 z/4 )}$, where $d_1,d_2$ are arbitrary constants.
\end{theorem}

{\bf Proof.}
Because of the symmetry (\ref{involut}) it is sufficient to study either the
equations (\ref{cubm}) or (\ref{cubp}) for $c\in \R$ or both these equations
for $c>0$. The value $c=0$ can be excluded because it complies with the
constant value of $p(z) $ and correspondingly constant value of $v_z(z)$, but
all such cases are studied before and the solutions are given by
(\ref{trivisol1}).

We will study equation (\ref{cubp}) in case $c \in \R \setminus \{0\}$ and
obtain on this way the complete class of exact solutions for equations
(\ref{cubm})-(\ref{cubp}) and on this way for the equation (\ref{speck1q4}).

Equation (\ref{cubp}) for $c>0$ has a one real root only. It leads to
an ordinary differential equation of the form
\begin{eqnarray}
 v_z(z)&=& \frac{1}{b}  \left( 1 +
   \left( 1 + c\,e^{-\frac{3\,z}{2}} +
          {\sqrt{2\,c\,e^{-\frac{3\,z}{2}} + c^2\,e^{-3\,z}}} \right)^
       {-\frac{2}{3}} \nonumber
 \right. \\
&& \left.  + \left( 1 + c\,e^{-\frac{3\,z}{2}} +
          {\sqrt{2\,c\,e^{-\frac{3\,z}{2}} + c^2\,e^{-3\,z}}} \right)^
       {\frac{2}{3}} \right)
, ~~c>0. \label{reshv}
\end{eqnarray}

Equation (\ref{reshv}) can be exactly integrated if we use an Euler
substitution and introduce a new independent variable
\begin{equation}
\tau=2 \left(1+ c\,e^{\frac{-3\,z}{2}} + {\sqrt{2\,c\,e^{\frac{-3\,z}{2}} +
       c^2\,e^{-3\,z}}}\right). \label{podstau1}
\end{equation}

The corresponding solution is given by
\begin{eqnarray}
v_r(z) &= -
     \frac{1}{b} \left( 1 + c\,e^{-\frac{3\,z}{2}} +
           {\sqrt{2\,c\,e^{-\frac{3\,z}{2}} + c^2\,e^{- 3\,z}}} \right)^
        {-\frac{2}{3}}  \label{solvzplus}\\
&-   \frac{1}{b}{\left( 1 + c\,e^{-\frac{3\,z}{2}} +
         {\sqrt{2\,c\,e^{-\frac{3\,z}{2}} + c^2\,e^{- 3\,z}}} \right) }^
      {\frac{2}{3}} \nonumber\\
&- \frac{ 2}{b} \log \left(
      {\left( 1 + c\,e^{-\frac{3\,z}{2}} +
           {\sqrt{2\,c\,e^{-\frac{3\,z}{2}} + c^2\,e^{- 3\,z}}} \right) }^{-\frac{1}{3}}
\right. \nonumber \\
 & ~~~~~ + \left. \left( 1 + c\,e^{-\frac{3\,z }{2}} +
         {\sqrt{2\,c\,e^{-\frac{3\,z}{2}} + c^2\,e^{- 3\,z}}} \right)^
      {\frac{1}{3}} -2 \right)  +d\,,\nonumber
\end{eqnarray}
where $d \in \R$ is an arbitrary constant. If in the right hand side of
equation (\ref{cubp}) the parameter $c $ satisfies the inequality $c<0$ and
the variable $z$ chosen in the region
\begin{equation}
 z \in \left(-\frac{2}{3}\ln{\frac{2}{|c|}},\infty \right)
\end{equation}
then the equation on $p$ possesses maximal three real roots. These three roots
of cubic equation (\ref{cubp}) give rise to three differential equations of
the type $v_z=-(1- p^2(z))/b$. The equations can be exactly solved and we find
correspondingly three solutions $v_i(z),~ i=1,2,3$.

\begin{figure}[ht]
\vspace{0.5cm}
\begin{center}
\begin{minipage}[t]{12cm}
\includegraphics[width=12cm]{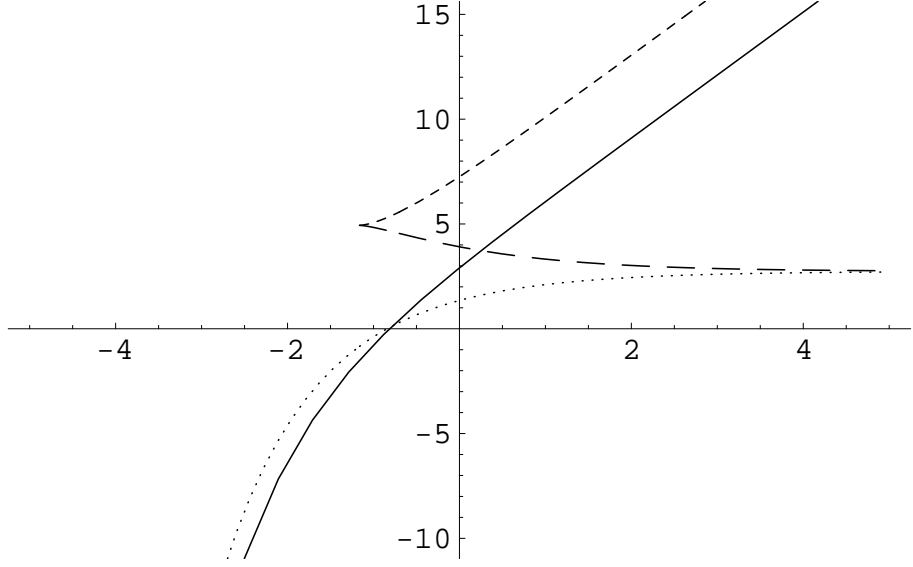}
\caption{\label{solfi2} Plot of the solution $v(z)$ given in (\ref{solvzplus})
(thick solid line); $v_1(z)$ from (\ref{resh1}) (short dashed line); $v_2(z)$
from (\ref{resh2}) (long dashed line) and the third solution
$v_{3,1}(z),v_{3,2}(z) $ from (\ref{resh31}), (\ref{resh32}) (thin solid
line). The parameters takes the values $~|c|=0.35,q=-4, d=0, b=1.$ and the
variable $z\in (-5,4.5)$. }
\end{minipage}
\end{center}
\end{figure}

The first solution is given by the expression
\begin{eqnarray}
v_1(z)  &=&-\frac{z}{b}-\frac{2}{b}\,\cos \left(\frac{2}{3}\,\arccos
\left(1 - |c|\,e^{\frac{-3\,z}{2}}\right)\right) \label{resh1}\\
&-& \frac{4}{3 b}\,\log \left(1 + 2\,\cos
\left(\frac{1}{3}\,\arccos
\left(1 - |c|\,e^{\frac{-3\,z}{2}}\right)\right)\right)\nonumber\\
&-& \frac{16}{3 b}
     \,\log \left(\sin
\left(\frac{1}{6}\,\arccos
\left(1 - |c|\,e^{\frac{-3\,z}{2}}\right)\right)\right)+d,\nonumber
\end{eqnarray}
where $d \in \R$ is an arbitrary constant. The second solution is given by the formula
\begin{eqnarray}
v_2(z) &=&-\frac{z}{b}-\frac{2}{b}\,\cos \left(\frac{2}{3} \pi -\frac{2}{3}\,\arccos
\left(1 - |c|\,e^{\frac{-3\,z}{2}}\right)\right) \label{resh2}\\
&-& \frac{4}{3 b}\,\log \left(-1 + 2\,\cos
\left(\frac{1}{3} \pi -\frac{1}{3}\,\arccos
\left(1 - |c|\,e^{\frac{-3\,z}{2}}\right)\right)\right)\nonumber\\
&-& \frac{16}{3 b}
     \,\log \left(\sin
\left(\frac{1}{6} \pi -\frac{1}{6}\,\arccos
\left(1 - |c|\,e^{\frac{-3\,z}{2}}\right)\right)\right)+d,\nonumber
\end{eqnarray}
where $d \in \R$ is an arbitrary constant. The first and second solutions are
defined up to the point $z=-\frac{2}{3}\ln{\frac{2}{|c|}}$ where they coincide
(see Fig.~\ref{solfi2}).

The third solution for $z> -\frac{2}{3}\ln{\frac{2}{|c|}}$ is given by the
formula
\begin{eqnarray}
v_{3,2}(z) &=&-\frac{z}{b} -\frac{2}{b}\,\cos \left(\frac{2}{3}\pi +\frac{2}{3}\,\arccos
\left(1 - |c|\,e^{\frac{-3\,z}{2}}\right)\right) \label{resh31}\\
&-& \frac{4}{3 b}\,\log \left(-1 + 2\,\cos
\left(\frac{1}{3}\pi +\frac{1}{3}\,\arccos
\left(-1 + |c|\,e^{\frac{-3\,z}{2}}\right)\right)\right)\nonumber\\
&-& \frac{16}{3 b}
     \,\log \left(\cos
\left(\frac{1}{6} \pi +\frac{1}{6}\,\arccos
\left(-1 + |c|\,e^{\frac{-3\,z}{2}}\right)\right)\right)+d,\nonumber
\end{eqnarray}
where $d \in \R$ is an arbitrary constant. In case $z<
\frac{2}{3}\ln{\frac{2}{|c|}}$ the polynomial (\ref{cubp}) has a one real root
and the corresponding solution can be represented by the formula
\begin{eqnarray}
v_{3,1}(z) &=& -\frac{z}{b} - \frac{2}{b}\,\cosh \left(\frac{2}{3}
          {\rm arccosh}\left(-1 +
          |c|\,e^{\frac{-3\,z}{2}}\right)\right)\label{resh32}\\
&-&
    \frac{16}{3\,b}\,\log \left(\cosh \left(\frac{1}{6}{\rm arccosh}\left(-1 +
            |c|\,e^{\frac{-3\,z}{2}}\right)\right)\right)\nonumber \\
&-&
   \frac{4}{3\,b}\,\log \left(-1 + 2\,\cosh \left(\frac{1} {3} {\rm arccosh} \left(-1 +
              |c|\,e^{\frac{-3\,z}{2}}\right)\right)\right) +d.\nonumber
\end{eqnarray}

The third solution is represented by formulas $v_{3,2}(z)$ and  $v_{3,1}(z)$
for different values of the variable $z$.
\qquad

One of the sets of solutions (\ref{solvzplus}), (\ref{resh1}) -(\ref{resh32})
for fixed parameters $b,c,d$ is represented in Fig.~\ref{solfi2}. The first
solution (\ref{reshv}) and the third solution given by both (\ref{resh31}) and
(\ref{resh32}) are defined for any values of $z$. The solutions $v_1(z)$ and
$v_2(z)$ cannot be continued to the left after the point
$z=-\frac{2}{3}\ln{\frac{2}{|c|}}$ where they coincide.

\begin{figure}[ht]
\vspace{0.5cm}
\begin{center}
\begin{minipage}[t]{12cm}
\includegraphics[width=12cm]{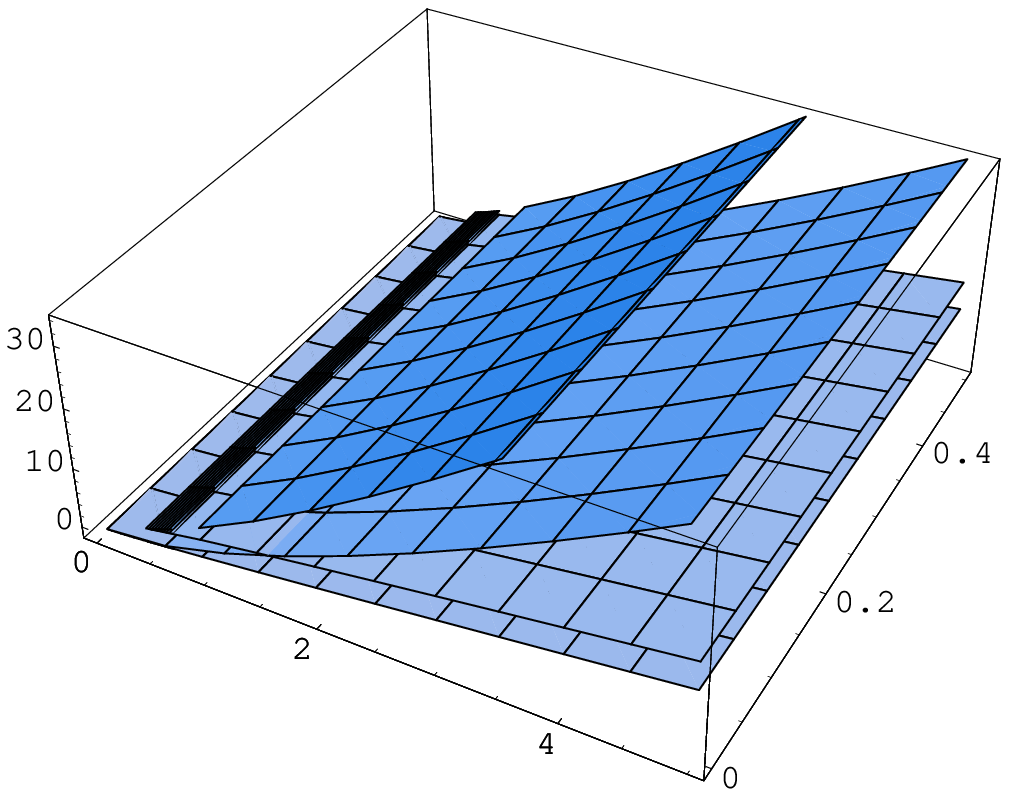}
\caption{\label{solfi3} Plot of solutions $u_r(S,t),u_1(S,t),u_2(S,t)$ and $u_{3,1}(S,t),u_{3,2}(S,t) $  for the
  parameters $\sigma=0.4,~|c|=0.5,q=-4,b=1.0, d=0$. The variables $S,t$ lie in intervals $S\in (0.,5.)$ and $t\in [0,0.5]$.
All invariant solutions change slowly in $t$-direction.}
\end{minipage}
\end{center}
\end{figure}

If we keep in mind that $z=\log S - \frac{\sigma^2}{8}t$ and $u(S,t)=S ~ v(z)$
we can represent the complete set of exact invariant solution of equation
(\ref{eq:nonlinear-BS-SDE}). The solution (\ref{solvzplus}) gives rise to an
invariant solution $u_r(S,t)$ in the form
\begin{eqnarray}
u_r(S,t) &=& - \frac{1}{\omega \rho} S \left( 1 + c\,S^{-\frac{3}{2}}
      e^{\frac{3 \sigma^2}{16} t} + {\sqrt{2\,c\,S^{-\frac{3}{2}} e^{\frac{3
      \sigma^2}{16}t } + c^2\,S^{-3} e^{\frac{3 \sigma^2}{8}t}}}
      \right)^{-\frac{2}{3}} \nonumber\\
&-& \frac{1}{\omega \rho} S {\left( 1
      + c\,S^{-\frac{3}{2}} e^{\frac{3 \sigma^2}{16}t} +
      {\sqrt{2\,c\,S^{-\frac{3}{2}} e^{\frac{3 \sigma^2}{16}t} + c^2\,S^{-3}
      e^{\frac{3 \sigma^2}{8}t}}} \right) }^
      {\frac{2}{3}}~~~~~\label{solustl}\\
&-& \frac{2}{\omega \rho} S \log
      \left( {\left( 1 + c\,S^{-\frac{3}{2}} e^{\frac{3 \sigma^2}{16}t} +
      {\sqrt{2\,c\,S^{-\frac{3}{2}} e^{\frac{3 \sigma^2}{16}t} + c^2\,S^{-3}
      e^{\frac{3 \sigma^2}{8}t}}} \right) }^ {-\frac{1}{3}}
      \right. \nonumber\\
&+& \left. {\left( 1 +
      c\,S^{-\frac{3}{2}} e^{\frac{3 \sigma^2}{16}t} +
      {\sqrt{2\,c\,S^{-\frac{3}{2}} e^{\frac{3 \sigma^2}{16}t} + c^2\,S^{-3}
      e^{\frac{3 \sigma^2}{8}t}}} \right) }^ {\frac{1}{3}} -2 \right) + d ~ S
      +d_2\nonumber
\end{eqnarray}
where $d ,~d_2\in \R$,  $c>0$. This solution was obtained and studied in
\cite{Bordag:2004}. We describe now other invariant solutions from the complete set of
invariant solutions.

In case $c<0$ we can obtain correspondingly three  real solutions if
\begin{equation} \label{ogranst}
 S\ge \left(\frac{|c|}{2}\right)^{2/3} \exp{\left(\frac{\sigma^2}{8} t \right)}.
\end{equation}
The first solution is represented by
\begin{eqnarray}
&&u_1(S,t) =-\frac{1}{\omega \rho} S \left(\log {S} - \frac{\sigma^2}{8}
t\right)-\frac{2}{\omega \rho}\, S \cos \left(\frac{2}{3}\,\arccos \left(1 -
|c|\, S^{-\frac{3}{2}} e^{\frac{3 \sigma^2}{16}t} \right)\right) \nonumber\\
&&- \frac{4}{3 \omega \rho}\, S \log \left(1 + 2\,\cos
\left(\frac{1}{3}\,\arccos
\left(1 - |c|\,
S^{-\frac{3}{2}} e^{\frac{3 \sigma^2}{16}t}
\right)\right)\right)\label{solustl1}\\
&&- \frac{16}{3 \omega \rho} \, S \log \left(\sin \left(\frac{1}{6}\,\arccos
\left(1 - |c|\,
S^{-\frac{3}{2}} e^{\frac{3 \sigma^2}{16}t}
\right)\right)\right)+d ~ S + d_2,\nonumber
\end{eqnarray}
where $d , ~ d_2\in \R$,  $c<0$. The second solution is given by the formula
\begin{eqnarray}
u_2(S,t) &=&-\frac{1}{\omega \rho} S \left( \log {S} - \frac{\sigma^2}{8}
t\right) -\frac{2}{\omega \rho}\, S \cos \left(\frac{2}{3} \pi
+\frac{2}{3}\,\arccos \left(-1 + |c|\, S^{-\frac{3}{2}} e^{\frac{3
    \sigma^2}{16}t} \right)\right) \nonumber\\ &-& \frac{4}{3 \omega \rho}\, S
\log \left(1 + 2\,\cos \left(\frac{1}{3} \pi +\frac{1}{3}\,\arccos \left(-1 +
|c|\, S^{-\frac{3}{2}} e^{\frac{3 \sigma^2}{16}t}\right)\right)\right)
\label{solustl2}\\ &-& \frac{16}{3 \omega \rho} \, S \log \left(\sin
\left(\frac{1}{6} \pi +\frac{1}{6}\,\arccos \left(-1 + |c|\, S^{-\frac{3}{2}}
e^{\frac{3 \sigma^2}{16}t}\right)\right)\right)+d ~ S +d_2. \nonumber
\end{eqnarray}
where $d , ~ d_2 \in \R$, $c<0$. The first and second solutions are defined
for the variables under conditions (\ref{ogranst}). They coincide along the
curve
$$S=\left(\frac{|c|}{2}\right)^{2/3} \exp{\left(\frac{\sigma^2}{8} t
\right)}$$ and cannot be continued further. The third solution is defined by
\begin{eqnarray}
u_{3,2}(S,t) &=&-\frac{1}{\omega \rho} S \left(\log {S} - \frac{\sigma^2}{8}  t\right) \label{solustl31}\\
&-& \frac{2}{\omega \rho}\, S \cos \left(\frac{2}{3}\,\arccos
\left(-1 + |c|\,
S^{-\frac{3}{2}} e^{\frac{3 \sigma^2}{16}t}
\right)\right) \nonumber\\
&-& \frac{4}{3 \omega \rho}\, S \log \left(-1 + 2\,\cos
\left(\frac{1}{3}\,\arccos
\left(-1
  + |c|\,
S^{-\frac{3}{2}} e^{\frac{3 \sigma^2}{16}t}
 \right)\right)\right) \nonumber\\
&-& \frac{16}{3 \omega \rho}
     \, S \log \left(\cos \left(\frac{1}{6}\,\arccos \left(-1 + |c|\,
S^{-\frac{3}{2}} e^{\frac{3 \sigma^2}{16}t}
\right)\right)\right)+d ~ S +d_2, \nonumber
\end{eqnarray}
where $d , ~ d_2\in \R$ and $S,t$ satisfied the condition (\ref{ogranst}).

In case $0<S<\left(\frac{|c|}{2}\right)^{ \frac{2}{3}}
\exp{\left(\frac{\sigma^2}{8} t \right)}$ the third solution can be
represented by the formula
\begin{eqnarray}
u_{3,1}(S,t)&=& -\frac{1}{\omega \rho} S \left(\log {S} - \frac{\sigma^2}{8}
t\right) \label{solustl32} \\ &-& \frac{2}{\omega \rho}\, S \cosh
\left(\frac{2}{3} {\rm arccosh}\left(-1 + |c|\, S^{-\frac{3}{2}} e^{\frac{3
\sigma^2}{16}t} \right)\right)\nonumber \\ &-& \frac{16}{3\,\omega \rho}\, S
\log \left(\cosh \left(\frac{1}{6}{\rm arccosh}\left(-1 + |c|\,
S^{-\frac{3}{2}} e^{\frac{3 \sigma^2}{16}t} \right)\right)\right) \nonumber\\
&-& \frac{4}{3\,\omega \rho}\, S \log \left(-1 + 2\,\cosh \left(\frac{1} {3}
{\rm arccosh} \left(-1 + |c|\, S^{-\frac{3}{2}} e^{\frac{3 \sigma^2}{16}t}
\right)\right)\right) \nonumber \\
&+& d ~ S +d_2,\nonumber
\end{eqnarray}
where $d , ~ d_2\in \R.$ The solution $u(S,t)$ (\ref{solustl}) and the third
solution given by $u_{3,1}$, $u_{3,2}$ (\ref{solustl31}),(\ref{solustl32}) are
defined for all values of variables $t$ and $S>0$. They have a common
intersection curve of the type $S={\rm const.}~ \exp({\frac{ \sigma^2}{8}t}
)$. The typical behavior of all these invariant solutions is represented in
 Fig.~\ref{solfi3}.

The previous results we  summarize in the following theorem which describes
the set of invariant solutions of equation (\ref{eq:nonlinear-BS-SDE}).

\begin{theorem}\label{maintheorem}

The invariant solutions of equation (\ref{eq:nonlinear-BS-SDE}) can be defined
by the set of first order ordinary differential equations
listed in Theorem~\ref{theoremq}.\\
 If moreover the
parameter $q=-4$, or equivalent in the substitution (\ref{newvar}) we chose
$a=-\sigma^2/8$ then the complete set of invariant solutions of
(\ref{eq:nonlinear-BS-SDE}) can be found exactly. This set of invariant
solutions is given by formulas (\ref{solustl})--(\ref{solustl32}) and by
solutions
\begin{eqnarray}
u(S,t)=d ~ S,~~u(S,t)=\frac{3}{b} ~ S ~(\log{S}- \sigma^2
\frac{t}{8}),\nonumber \\
~~u(S,t)=- \frac{1}{b} ~ S ~(\log{S}-\sigma^2
 \frac{ t}{8}), \nonumber
\end{eqnarray}
where $d$ is an arbitrary constant. This set of invariant solutions is unique
up to the transformations of the symmetry group $G_{\Delta}$ given by Theorem
\ref{symteor}.
\end{theorem}

The solutions $u_r(S,t)$, (\ref{solustl}), $u_1(S,t)$, (\ref{solustl1}),
$u_2(S,t)$, (\ref{solustl2}), $u_{3,1}(S,t)$ ,(\ref{solustl31}),
$u_{3,2}(S,t)$ (\ref{solustl32}), have no one counterpart in a linear case.
If in the equation (\ref{eq:nonlinear-BS-SDE}) the parameter $\rho=0$
equation becomes linear. If the parameter $\rho \to 0$ then equation
(\ref{eq:nonlinear-BS-SDE}) and correspondingly equation (\ref{uravlam}) will
be reduced to the linear Black-Scholes equation but solutions
(\ref{solustl})-(\ref{solustl32}) which we obtained here will be completely
blown up by $\rho \to 0$ because of the factor $1/b=1/(\omega \rho)$ in the
formulas (\ref{solustl})-(\ref{solustl32}).  This phenomena was described as
well in \cite{Bordag:2004}, \cite{Chmakova} where the solution $u_r(S,t)$ was
studied and for the complete set of invariant solutions of equation
(\ref{uravlam}) with $k=1$ in \cite{Bordag:2005}.

\section{Properties of solutions and parameter-sensitivity}

We study the properties of solutions, keeping in mind that because of the
symmetry properties (see Theorem~\ref{symteor}) of the equation
(\ref{eq:nonlinear-BS-SDE}) we can add to each solution a linear function of
$S$.

\subsection{Dependence on the constant $c$ and terminal payoff}

\begin{figure}[ht]
\vspace{0.5cm}
\begin{minipage}[t]{6.5cm}
\includegraphics[width=6.5cm]{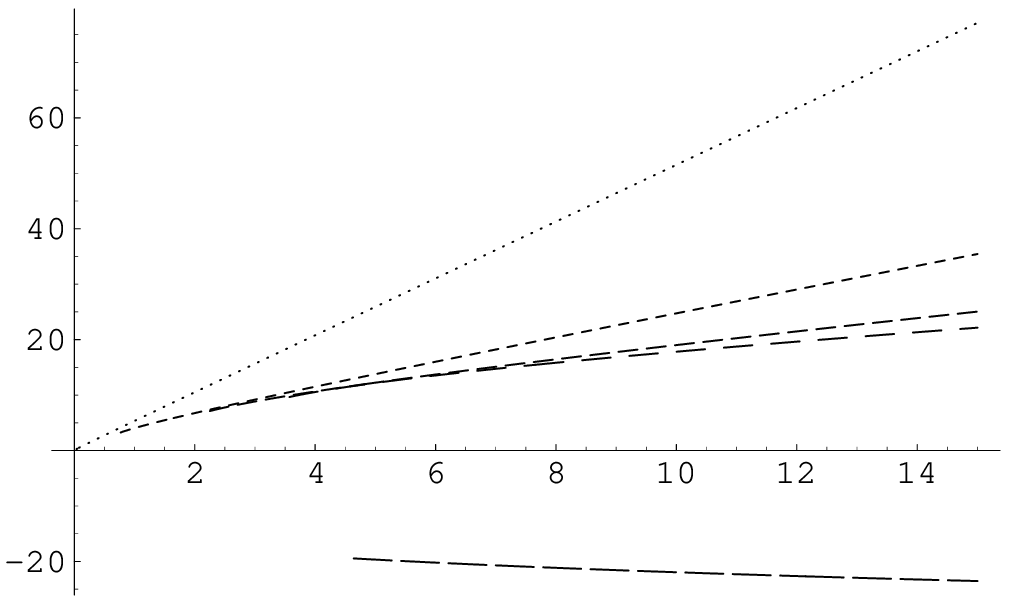}
\caption{\label{depcu2} Plot of solutions $u_2(S,t) $ where
  $~|c|=0.01,1.,5.,10.,20.,\sigma=0.4, q=-4$,
$ b=1.0, d=0$. The
  variables $S$ lie in intervals $S\in (0.,15.)$ and $t=1.$.}
The curves going from up to down with the growing value of the parameter $|c|$.
\end{minipage} \hfill
\begin{minipage}[t]{6.5cm}
\includegraphics[width=6.5cm]{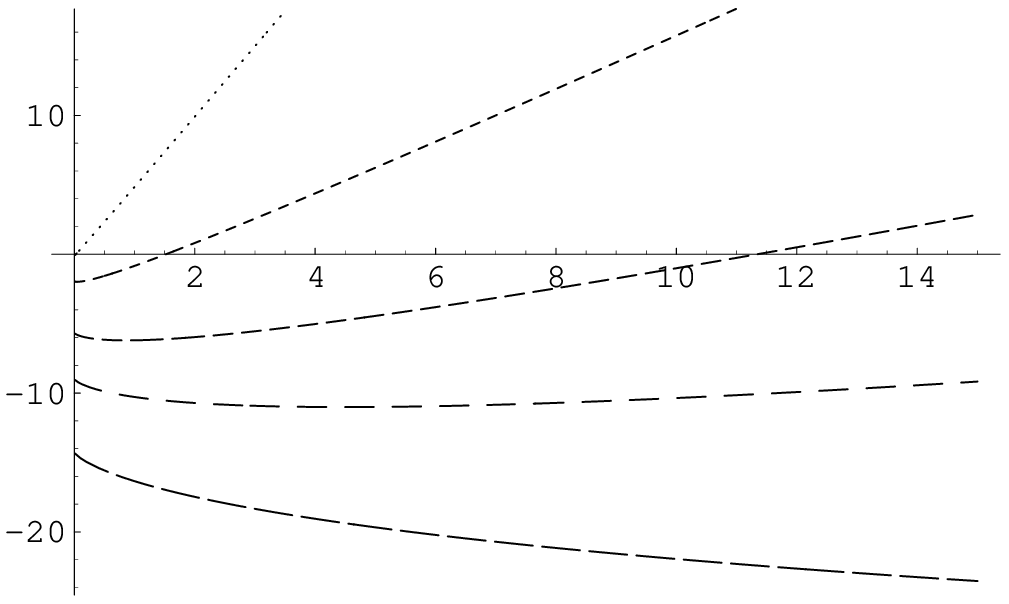}
\caption{\label{depcu3} Plot of solutions $u_3(S,t) $ where
  $|c|=0.01,1.,5.,10.,20., \sigma=0.4, q=-4$,
$ b=1.0, d=0$. The
  variables $S$ lie in intervals $S\in (0.,15.)$ and $t=1.$.}
The curves going from up to down with the growing value of the parameter $|c|$.
\end{minipage}
\end{figure}

First we study the dependence of the solution $u_{2}$ on the arbitrary
constant $c$. The constant $c$ is the first constant of integration of the
ordinary differential equations (\ref{yzeqm}), (\ref{yzeqp}). This dependence
is illustrated in Figure~\ref{depcu2}.
We see on Fig.~\ref{depcu2} that for different values of the constant $c$ the
domains of the solutions are different.  The dependence of the solutions
$u_{3,1},u_{3,2}(S,t)$ on the constant $c$ is exemplified in
Figure~\ref{depcu3}.

We obtain a typical terminal payoff function for the solutions
(\ref{solustl})-(\ref{solustl32}) if we just fix $t=T$. By changing the value
of the constant $c$ and by adding a linear function of $S$ we are able to
modify terminal payoff function for the solutions. Hence we can approximate
typical payoff profiles of financial derivatives quite well.

\subsection{Dependence on time}

All solutions depend weakly on time because of the substitution (\ref{podszotst}) all
invariant solutions depends on the combination $\sigma^2 t$. As long as  we take the
volatility $\sigma$ to be small we obtain a dampened dependence of the solutions on
time.

\begin{figure}[ht]
\vspace{0.5cm}
\begin{minipage}[t]{12cm}
\includegraphics[width=12cm]{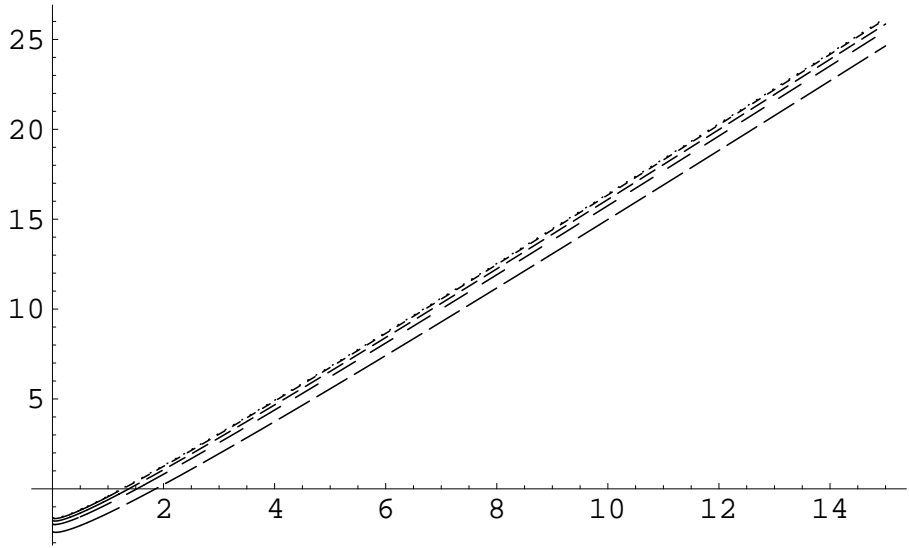}
\caption{\label{dept} Plot of solutions $u_{3,1}(S,t),  u_{3,2}(S,t)$  for the
  parameters $\sigma=0.4,~|c|=10.,q=-4,b=1.0, d=0$.
The variables $S$ lie in intervals $S\in (0.,15.)$ and
  $t=0.01,1.,5.,10.,20.$.}
The highest level corresponds to $t=0.01$ and the lowest to $t=20.$.
\end{minipage}
\end{figure}

In Fig.~\ref{dept} we can see this dependence for  examples of the solution
$u_{3,1}(S,t), u_{3,2}(S,t)$.

\subsection{Dependence on the parameter $\rho$}\label{rhosav}

All solutions found in this paper have the form
\begin{equation}
u(S,t)=w(S,t)/\rho, \label{symprho}
\end{equation}
where $w(S,t)$ is a smooth function of $S,t$. Hence the function $w(S,t)$ solves the
equation (\ref{eq:nonlinear-BS-SDE}) with $\rho=1$,
\begin{eqnarray} \label{w}
w_t+\frac{\sigma^2 S^2}2\frac{w_{SS}}{(1- S w_{SS})^2}=0. \label{nonl}
\end{eqnarray}
Because of this relation any $\rho$-dependence of invariant solutions of
(\ref{eq:nonlinear-BS-SDE}) trivial.  In particular, if the terminal
conditions are fixed, $u(S,T)=h(S)$, then the value $u(S,t)$ will increase if
the value of the parameter $\rho$ increases. This dependence of hedge costs on
the position of the large trader on the market is very natural.

\subsection{Dependence on the  asset price $S$}

\begin{figure}[ht]
\vspace{0.5cm}
\begin{minipage}[t]{6.5cm}
\includegraphics[width=6.5cm]{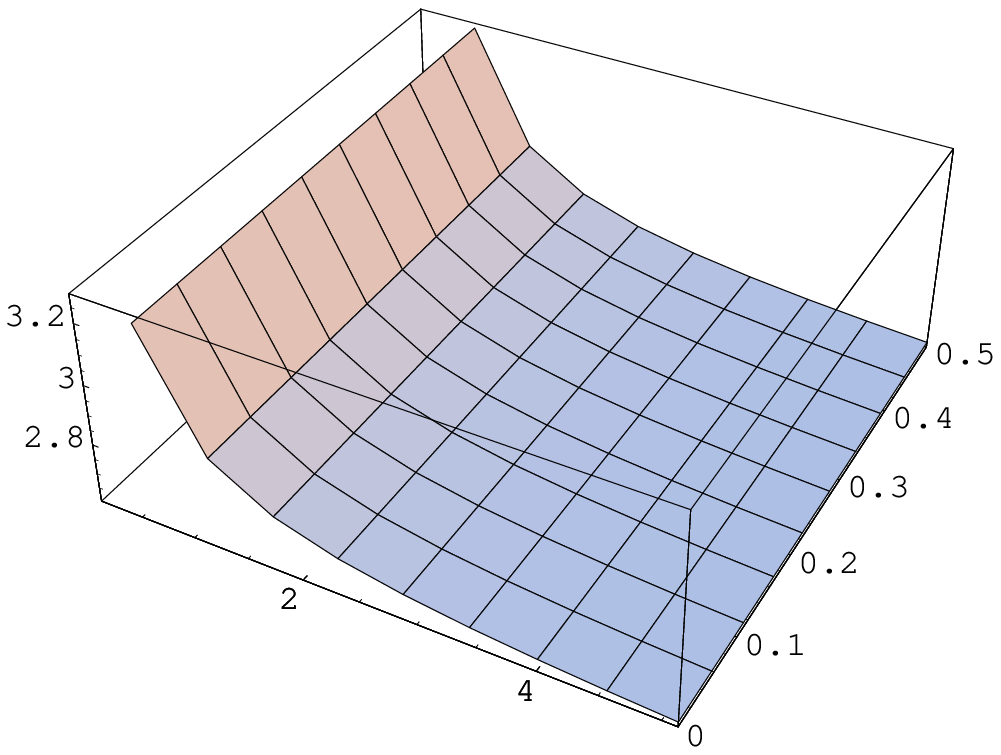}
\caption{\label{deltau2} Plot of $\Delta$ for $u_2(S,t)$ and
$\sigma=0.4,~|c|=1.,q=-4,
b=1.0, d=0$. The variables lie in intervals $S\in
(0.,5.)$ and $t=[0.,0.5]$.}
\end{minipage}  \hfill
\begin{minipage}[t]{6.5cm}
\includegraphics[width=6.5cm]{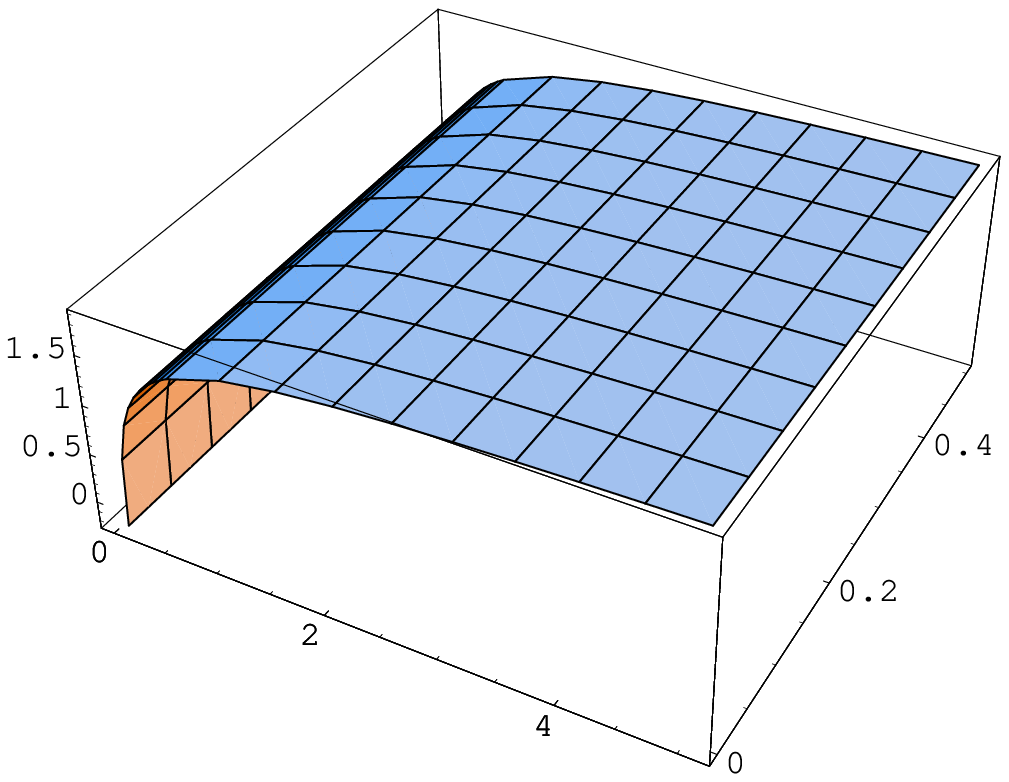}
\caption{\label{deltau3} Plot of the $\Delta$ for $u_{3,1}(S,t),
  u_{3,2}(S,t) $ and $\sigma=0.4,~|c|=1.,q=-4,
b=1.0, d=0$.
The variables lie in intervals $S\in (0.,5.)$ and $t\in [0.01,0.5]$.}
\end{minipage}
\end{figure}

In practice one use often delta-hedging to reduce the sensitivity of a
portfolio to the movements of an underlying asset. Hence it is important to
know the value $\Delta$ defined by
$
\Delta=\frac{\partial u }{\partial S},
$
where $u$ denotes the value of the derivative product or of a portfolio. Using
the exact formulas for the invariant solutions we can easily calculate
$\Delta$ as a function of $S$ and $t$. The $\Delta$ corresponding to the
solution $u_2$ is presented in Fig.~\ref{deltau2}

The $\Delta$ of the solution $u_{3,1},u_{3,2}$ is represented on the
Fig.~\ref{deltau3}.\\  We see in both cases the strong dependence on $S$ for small value
of $S$. If $S \to \infty $, in both cases the $\Delta$ tends to a constant which is
independent of time and the constant $c$.

\subsection{Asymptotic behavior of  invariant solutions}

If $S$ is large enough we have four well defined solutions $u_r(S,t)$, $u_1(S,t)$,
$u_2(S,t)$, $u_{3,2}(S,t)$.  The asymptotic behavior solutions $u_r(S,t)$ from
(\ref{solustl}) and $u_{1}$ from (\ref{solustl1}) coincides in the main terms as $S
\to \infty$ and is given by the formula
\begin{equation}
u_r(S,t),~u_{1}(S,t) \sim \frac{1}{b} \left(  3 S \ln S + {\rm const.} S
+{\cal O} (S^{-1/2}) \right) \label{asymr1}
\end{equation}
The exact formula for the asymptotic behavior of the function $u_1(S,t)$ for
$S \to \infty $ is given by
\begin{eqnarray}
u_{1}(S,t) &\sim& \frac{3}{b}\,S \log (S)+
\frac{1}{b} S \left(
4\,\log (3)- 2  - \frac{8}{3}\,\log \left(\frac{2}{|c|}\right) - \frac{3}{8} \sigma^2\,t
\right) \nonumber \\
 &+&
   \frac{2^4}{3^3 b}\,|c|\,e^{\frac{3\,{\sigma}^2\,t}{16}}\,S^{-\frac{1}{2}}
     + {\cal O}\left(S^{-\frac{5}{4}}\right). \label{asymu1infty}
\end{eqnarray}
We see that for $S \to \infty $, the main term does not depend on time or on
the value of the constant $c$. Moreover, this term cannot be changed by adding
a linear function of $S$ to the solution.

The main terms of the solutions $u_2(S,t)$ from (\ref{solustl2}) and  $u_{3,2}$ from
(\ref{solustl32}) behave similar to each other as $S \to \infty$; this behavior is
given by the formulas
\begin{eqnarray}
u_{2}(S,t) &\sim& \frac{1}{b} \left( 1 + \frac{2}{3}\,\log
\left(\frac{2^7}{3^3 |c|}\right) \right) \,S \label{asymu2infty}\\
&+&\frac{2^3}{3
b}\,\sqrt{\frac{2|c|}{3}}\,e^{\frac{3\,{\sigma}^2\,t}{32}}\,S^{\frac{1}{4}} -
\frac{2^3}{3^3 b}\,|c|\,e^{\frac{3\,{\sigma}^2\,t}{16}}\,S^{-\frac{1}{2}} +
{\cal O}\left(S^{-\frac{5}{4}}\right). \nonumber
\end{eqnarray}
and
\begin{eqnarray}
u_{3,2}(S,t) &\sim& \frac{1}{b} \left( 1 + \frac{2}{3}\,\log
\left(\frac{2^7}{3^3 |c|}\right) \right) \,S \label{asymu3infty} \\
&-&\frac{2^3}{3
b}\,\sqrt{\frac{2|c|}{3}}\,e^{\frac{3\,{\sigma}^2\,t}{32}}\,S^{\frac{1}{4}} -
\frac{2^3}{3^3 b}\,|c|\,e^{\frac{3\,{\sigma}^2\,t}{16}}\,S^{-\frac{1}{2}} +
{\cal O}\left(S^{-\frac{5}{4}}\right).
\end{eqnarray}
Note that the main term in formulas (\ref{asymu2infty}) and (\ref{asymu3infty})
depends on $S$ linearly and has a coefficient which depends on the constant $c$ only.
Hence we can change the asymptotic behavior of the solutions $u_2(S,t)$ and
$u_{3,1}(S,t)$, $u_{3,2}(S,t)$ by simply adding a linear function of $S$ to the
solutions.

For $S$ in a neighborhood of $S \to 0$ there exist just two non trivial
real invariant
solutions of equation (\ref{eq:nonlinear-BS-SDE}), i.e. the known solution
$u_r(S,1)$ and the new solution $u_{3,1}(S,t)$. Using the exact formulas
for the last  solution we retain the first term and obtain as $S \to 0$ for the
solution $u_{3,1}(S,t)$
\begin{eqnarray}
u_{3,1}(S,t) \sim \frac{-14}{b}~ S \ln \left(S\right)
 +{\cal O}\left(S\right) .
\end{eqnarray}
The main term in this formula do not depend on $t$ or constant $c$ and can
 be changed by adding a linear function of $S$ to the solution.

\section{Conclusion}

In this paper we have obtained explicit solutions for the equation
(\ref{eq:nonlinear-BS-SDE}) and studied their analytical properties. These
solutions are useful for a number of reasons. To begin with, while the payoffs
of these similarity solutions cannot be chosen arbitrary, the payoffs can be
modified using embedded constants to tailor a given portfolio reasonably
well. For some values of the parameters $c,d$ we obtain a payoff typical for
futures, for other values $c,d$ the payoff is very similar to the form of
calls (see Figure~\ref{depcu2} and Figure~\ref{depcu3}). Moreover, the
explicit solutions can be used as benchmark for different numerical methods.

An important difference between the case of the linear Black-Scholes equation
and these nonlinear cases can be noticed if we consider the asymptotic for $S
\to \infty $.  In the linear case the price of a Call option satisfies $u(S,t)
\to {\rm const} \cdot S $. In the nonlinear case the families of similarity
solutions $u_2$ and $u_3$ which approximate the payoff of a Call option well
on a finite interval $[0 , \bar{S}]$, grow faster than linear as $S \to
\infty$; see the formulas (\ref{asymu2infty}) and (\ref{asymu3infty}) for
details. This reflects the fact that in illiquid markets option hedging is
more expensive than in the standard case of perfectly liquid markets.

\section{Acknowledgements}

The authors are grateful to Albert N. Shiryaev for the interesting and
fruitful discussions.

\end{document}